\documentclass[english]{article}
\usepackage[T1]{fontenc}
\usepackage[latin9]{inputenc}
\usepackage{float}
\usepackage{amsmath}
\usepackage{amssymb}
\usepackage{graphicx}
\usepackage{esint}

\makeatletter

\floatstyle{ruled}
\newfloat{algorithm}{tbp}{loa}
\providecommand{\algorithmname}{Algorithm}
\floatname{algorithm}{\protect\algorithmname}

\usepackage{spconf,amsmath,graphicx,amsthm,amssymb,psfrag}
\usepackage{bibspacing}
\name{Andreas M\"{u}ller$^{\star}$ \qquad Dino Sejdinovic$^{\dagger}$ \qquad Robert Piechocki$^{\star}$}

\address{$^{\star}$ Merchant Venturers School of Engineering, University of Bristol, Bristol, BS8 1UB, UK\\
    $^{\dagger}$Gatsby Computational Neuroscience Unit, University College London, London, WC1N 3AR, UK\\ \{andreas.muller, r.j.piechocki\}@bristol.ac.uk, dino@gatsby.ucl.ac.uk}

\newcommand{\captionfonts}{\footnotesize}
\makeatletter  
\long\def\@makecaption#1#2{%
  \vskip\abovecaptionskip
  \sbox\@tempboxa{{\captionfonts #1: #2}}%
  \ifdim \wd\@tempboxa >\hsize
    {\captionfonts #1: #2\par}
  \else
    \hbox to\hsize{\hfil\box\@tempboxa\hfil}%
  \fi
  \vskip\belowcaptionskip}
\makeatother   

\makeatother

\usepackage{babel}
\begin{document}

\title{Approximate Message Passing under Finite Alphabet Constraints}
\maketitle
\begin{abstract}
In this paper we consider \emph{Basis Pursuit De-Noising} (BPDN) problems
in which the sparse original signal is drawn from a finite alphabet.
To solve this problem we propose an iterative message passing algorithm,
which capitalises not only on the sparsity but by means of a prior
distribution also on the discrete nature of the original signal. In
our numerical experiments we test this algorithm in combination with
a Rademacher measurement matrix and a measurement matrix derived from
the random demodulator, which enables compressive sampling of analogue
signals. Our results show in both cases significant performance gains
over a linear programming based approach to the considered BPDN problem.
We also compare the proposed algorithm to a similar message passing
based algorithm without prior knowledge and observe an even larger
performance improvement. 
\end{abstract}
\begin{keywords} Compressive Sampling, Signal Recovery, Finite Alphabet,
Message Passing\vspace{-0.1cm} \end{keywords}

\section{Introduction}

When the information bearing part of a signal lies only in a small
sub-space of the entire signal space, a uniform full rate sampling
approach, i.e., sampling at Nyquist rate, is inefficient. In \cite{Donoho2006}
and \cite{Candes2006}, Donoho and Candès \textit{et al.} addressed
this observation and introduced \emph{Compressive Sampling} (CS).
The key contribution of these papers was to show how random matrices
and $l_{1}$-minimization can be applied to achieve optimal recovery
of a sparse signal from a very limited number of measurements.

While initial reconstruction algorithms were based on convex optimisation
and linear programming, recent advances in compressive sampling have
led to the development of various other algorithms which solve the
reconstruction task at a lower computational complexity. Inspired
by the success of message passing algorithms as used for the decoding
operation of some channel codes, the authors of \cite{Baron2010}
solve the reconstruction of the original signal by means of belief
propagation on a sparse graph. More recently, a simple iterative soft
thresholding reconstruction algorithm, called \emph{Approximate Message
Passing} (AMP), has been proposed in \cite{Donoho2009}. The basis
of this algorithm is also belief propagation, albeit on a fully connected
graph, and it exhibits a virtually equivalent sparsity-undersampling
trade-off to that of linear programming based reconstruction algorithms.

In this paper, we address the problem of reconstructing a sparse finite
alphabet signal from a limited number of noisy measurements which
are obtained by compressive sampling. This problem appears in many
areas such as spectrum sensing, symbol detection in digital communications,
and multi-user detection, cf. \cite{Zhu2009,Tian2009}. Many existing
reconstruction algorithms for compressive sampling exploit the knowledge
of the sparsity level of the original signal. Building upon the AMP
framework, we propose here a novel AMP-based algorithm which capitalizes
not only on the sparsity but also on a prior distribution, which manifests
the finite alphabet property of the original signal.

In the derivation of this algorithm we assume that a time discrete
sparse signal is applied directly to a measurement matrix whose entries
are randomly sampled from $\left\{ -\frac{1}{\sqrt{R}},+\frac{1}{\sqrt{R}}\right\} $,
where $R$ is the number of measurements. In practice however, we
are typically confronted with various types of measurement matrices.
To this end we consider here as an application also the random demodulator
\cite{Tropp2010}, which allows to sample time-continuous analogue
signals at sub-Nyquist rates and resorts to techniques from compressive
sampling to reconstruct the signal. In both cases our numerical experiments
indicate that the proposed algorithm offers an excellent performance
in reconstructing sparse signals from noisy undersampled observations.
When we compare the new algorithm with a standard linear programming
based reconstruction algorithm and the AMP algorithm for the BPDN
problem from \cite{Donoho2010}, we observe a significant performance
improvement. Moreover, we note that the computational cost of the
algorithm proposed in this paper is comparable to existing state of
the art algorithms.

The remainder of the paper is organized as follows. In Section \ref{sec:Problem Outline}
we outline the considered compressive sampling problem. Next, we introduce
our proposed approximate message passing reconstruction algorithm
with prior knowledge in Section \ref{sec:Approximate Message Passing Algorithm with Discrete Prior Distribution}.
In Section \ref{sec:Rademacher Measurement Matrix} we compare the
proposed algorithm at first with the linear programming based SPGL1
algorithm and the AMP algorithm for the BPDN problem in combination
with a Rademacher measurement matrix. Then we briefly recap in Section
\ref{sec:Random Demodulator} on the random demodulator and also present
results for this scenario. Finally, Section \ref{sec:Conclusions}
concludes the paper.\vspace{-0.3cm}

\section{Problem Outline}

\label{sec:Problem Outline} Let $\mathcal{A}=\left\{ a_{1},\ldots,a_{S}\right\} $
denote a finite set of $S$ non-zero real numbers and define $\mathcal{A}_{0}:=\mathcal{A}\cup\left\{ 0\right\} $.
The $W$-dimensional sparse column vector $\mathbf{b}$ shall have
only $K\ll W$ non-zero entries, which are drawn uniformly at random
from $\mathcal{A}$. Moreover, let $\mathbf{\Psi}\in\mathbb{R}^{R\times W}$
be a matrix, for which $R\leq W$ holds. Given the noisy observation\vspace{-0.1cm}
\begin{equation}
\mathbf{v}=\mathbf{\Psi}\mathbf{b}+\mathbf{n},\label{eq:compressedNoisyMeasurement}
\end{equation}
where $\mathbf{n}=\left[n_{0},\ldots,n_{r},\ldots,n_{R-1}\right]^{\mathrm{T}}$
and $n_{r}$ is i.i.d. $\mathcal{N}\left(0,\sigma^{2}\right)$, our
aim is to recover $\mathbf{b}$. Problems of this type are frequently
considered in the CS literature. One way to reconstruct the sparse
vector $\mathbf{b}$ is by solving the optimization problem 
\begin{equation}
\arg\!\min_{\hat{\mathbf{b}}\in\mathcal{A}_{0}^{W}}\ \left\Vert \hat{\mathbf{b}}\right\Vert _{0}\ {\rm {subject\ to}\ \left\Vert \mathbf{\Psi}\hat{\mathbf{b}}-\mathbf{v}\right\Vert _{2}^{2}\leq\gamma',}\label{eq:lZeroNormOptimizationProblem}
\end{equation}
for an optimization constant $\gamma'$, which is chosen depending
on the noise variance. In general though the $l_{0}$-minimization
used in \eqref{eq:lZeroNormOptimizationProblem} is NP-hard. This
has led to various approximate algorithms, including those based on
convex relaxation, i.e., on $l_{1}$-minimization, like BPDN \cite{Chen2001}
or LASSO \cite{Tibshirani96}. Many of the approximate algorithms
for CS can thus be applied directly to approximate the solution of
\eqref{eq:lZeroNormOptimizationProblem} by solving the relaxed problem
\begin{equation}
\arg\!\min_{\hat{\mathbf{b}}\in\mathbb{R}^{W}}\ \left\Vert \hat{\mathbf{b}}\right\Vert _{1}\ {\rm {subject\ to}\ \left\Vert \mathbf{\Psi}\hat{\mathbf{b}}-\mathbf{v}\right\Vert _{2}^{2}\leq\gamma,}\label{eq:BpdnOptimizationProblem}
\end{equation}
and ensuring by, for example, thresholding that $\hat{\mathbf{b}}\in\mathcal{A}_{0}^{W}$
holds. However, in addition to the number of non-zero entries in the
sparse $W$-dimensional signal $\mathbf{b}$, the knowledge that its
entries lie in the set $\mathcal{A}_{0}$ can also be utilized directly
to improve the recovery of $\mathbf{b}$, as will be discussed in
the following sections.


\section{AMP with Discrete Prior Distribution}

\label{sec:Approximate Message Passing Algorithm with Discrete Prior Distribution}

It is well known that the solution of the $l_{1}$-minimization in
\eqref{eq:BpdnOptimizationProblem} corresponds to a mode of the posterior
distribution when a double-exponential prior distribution is used.
Donoho \textit{et al.} also use a double-exponential prior in the
derivation of their AMP algorithm \cite{Donoho2009}, where the contribution
of the prior distribution in the message update rules of the belief
propagation algorithm can be interpreted as a `sparsity promoting'
soft thresholding operation. However, their results are quite general
and a similar approach can be applied when a different choice of prior
distribution is more suitable. Indeed, \cite{Donoho2010} argues that
an estimate of the input distribution can be used to improve the recovery
algorithm. Therefore, it is natural to employ a discrete prior distribution
when the original signal is drawn from a finite alphabet, and we take
\begin{align}
f\left(b_{w}\right) & =\pi_{0}\cdot\delta\left\{ b_{w}=0\right\} +\sum_{s=1}^{S}\pi_{s}\cdot\delta\left\{ b_{w}=a_{s}\right\} \label{eq:discrete_prior}
\end{align}
as prior for each $b_{w}$ in $\mathbf{b}=\left[b_{0},...,b_{w},...,b_{W-1}\right]^{\mathrm{T}}$,
where $\pi_{0}=1-\frac{K}{W}$, and $\pi_{s}=\frac{K}{WS}$ for $1\leq s\leq S$.
Note that this prior is constructed under the assumption that the
non-zero entries in $\mathbf{b}$ are drawn uniformly from $\mathcal{A}$.
Should some non-zero entries be more likely than others, the prior
distribution can be easily modified to reflect this additional information.

Consider a fully connected bipartite graph between $R$ measurement
nodes on one side and $W$ variable nodes on the other side. The measurement
nodes shall represent the entries in $\mathbf{v}$ and likewise the
variable nodes the unknown entries in $\mathbf{b}$. As in \cite{Donoho2009}
we study belief propagation (BP) message updates between the measurement
and variable nodes on this complete graph. We denote the set of the
$W$ variable nodes as $\left[W\right]$ and use $w,\ \omega\in\left\{ 0,1,...,W-1\right\} $
as indices for this set. Similarly, we apply $r,\ \rho\in\left\{ 0,1,...,R-1\right\} $
as indices for the set $\left[R\right]$ of all $R$ measurement nodes.
At iteration $t$ we denote the message passed from the variable node
$w$ to the measurement node $r$ by $\nu_{w\to r}^{\left(t\right)}\left(b_{w}\right)$
and the message passed on this edge in the opposite direction by $\hat{\nu}_{r\to w}^{\left(t\right)}\left(b_{w}\right)$.
In the $t$-th iteration the BP message updates are then given as

\begin{eqnarray}
\hat{\nu}_{r\to w}^{\left(t\right)}\left(b_{w}\right) & \propto & \int\exp\left[-\frac{1}{2\sigma^{2}}(v_{r}-\left(\mathbf{\Psi b}\right)_{r})^{2}\right]\nonumber \\
 &  & \cdot\prod_{\omega\neq w}\nu_{\omega\to r}^{\left(t-1\right)}\left(b_{\omega}\right)d\mathbf{b}_{-w},\label{eq:factor to variable update}\\
\nu_{w\to r}^{\left(t\right)}\left(b_{w}\right) & \propto & f(b_{w})\prod_{\rho\neq r}\hat{\nu}_{\rho\to w}^{\left(t\right)}\left(b_{w}\right),\label{eq:variable to factor update}
\end{eqnarray}
where $d\mathbf{b}_{-w}$ denotes that integration is over all variables
except $b_{w}$, and $\sigma^{2}$ is the variance of the noise in
the measurements. Note that the messages from variable to measurement
nodes in \eqref{eq:variable to factor update} are proportional to
the probability mass function on $\mathcal{A}_{0}$. We initialize
\begin{align}
\nu_{w\to r}^{\left(0\right)}\left(b_{w}\right) & =f(b_{w}).\label{eq:variable to factor init}
\end{align}
We denote the mean and the variance of the message in \eqref{eq:variable to factor update}
by $\xi_{w\to r}^{\left(t\right)}$ and $\tau_{w\to r}^{\left(t\right)}$,
respectively. Note that $\xi_{w\to r}^{\left(0\right)}$ and $\tau_{w\to r}^{\left(0\right)}$
are initialized to the prior mean and variance $\forall w\in\left[W\right],\forall r\in\left[R\right]$.
We will derive the algorithm here under the assumption that $\psi_{r,w}\in\left\{ -\frac{1}{\sqrt{R}},+\frac{1}{\sqrt{R}}\right\} $.
Later in our simulations however we relax this assumption and work
with a general $\mathbf{\Psi}$ with normalized columns. 

Consider the random vector $\mathbf{b}_{-w}$$=[b_{0},b_{1},\ldots,b_{w-1},$
$b_{w+1},\ldots,b_{W-1}]^{\mathrm{T}}$ distributed according to the
product measure $\prod_{\omega\neq w}\nu_{\omega\to r}^{\left(t\right)}\left(b_{\omega}\right)$,
and the associated scalar random variable 
\begin{equation}
x_{r\to w}^{\left(t\right)}=v_{r}-\sum_{\omega\neq w}\psi_{r,\omega}b_{\omega}.
\end{equation}
Denote the induced density of $x_{r\to w}^{\left(t\right)}$ by $g$.
By the central limit theorem, for large $W$, $g$ can be approximated
by a Gaussian density with mean $v_{r}-\sum_{\omega\neq w}\psi_{r,\omega}\xi_{\omega\to r}^{\left(t\right)}$
and variance $\frac{1}{R}\sum_{\omega\neq w}\tau_{\omega\to r}^{\left(t\right)}$.
Thereby, if we write \eqref{eq:factor to variable update} in its
equivalent form
\begin{eqnarray*}
\hat{\nu}_{r\to w}^{\left(t\right)}\left(b_{w}\right) & \propto & \mathbb{E}_{x_{r\to w}^{(t)}\sim g}\exp\left[-\frac{1}{2\sigma^{2}}(x_{r\to w}^{(t)}-\psi_{r,w}b_{w})^{2}\right],
\end{eqnarray*}
we see that the factor to variable message update \eqref{eq:factor to variable update}
can be approximated by a Gaussian integral. These observations lead
to the simplified algorithm for message passing with a discrete prior
distribution given in Algorithm \ref{alg: AMP-with-discrete}%
\footnote{Throughout the algorithm, {}``dot'' is a placeholder for any $r\in\left[R\right]$
or for $\textrm{ø}$, in which case the summation in Step (3) is over
all $\rho\in\left[R\right]$.%
}. Note that we track the posterior probabilities in the log-domain,
as numerical simulations indicate that this results in a numerically
more stable implementation of the simplified message-passing algorithm.\vspace{-0.2cm}

\section{Numerical Experiments}

\begin{figure}
\vspace{-0.3cm}\centering \psfrag{detectionErrorProbability}[b][b]{$\mathrm{P}\left(\hat{b}_{w}\ne b_{w}\right)$}
\psfrag{-10Log10(SigmaSquared)}[t][t]{$-10\cdot\log_{10}\left(\sigma^{2}\right)$}
\includegraphics[width=1\columnwidth]{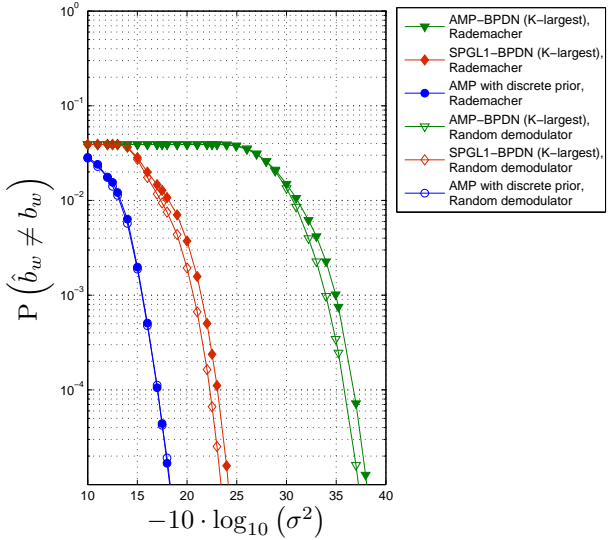}

\caption{Detection error rate $\mathrm{P}\left(\hat{b}_{w}\ne b_{w}\right)$
versus the noise variance $\sigma^{2}$ when the measurement matrix
$\Psi$ in \eqref{eq:compressedNoisyMeasurement} is a Rademacher
matrix with unit column norm, or the random demodulator matrix. The
number of iterations for both AMP algorithms is set to $T=50$.}

\label{fig:Fig4PaperRademacherDetectionErrorProbVsVariance_W=00003D512_R=00003D205_K=00003D20} 
\end{figure}
\label{sec:Numerical Experiments}

\subsection{Rademacher Measurement Matrix}

\label{sec:Rademacher Measurement Matrix} For the numerical experiments
presented here we assume that $\mathcal{A}_{0}=\{-1,0,+1\}$ and fix
$W=512$, $R=205$, $K=20$. The entries in the measurement matrix
$\mathbf{\Psi}$ shall be drawn uniformly at random from $\left\{ -1/\sqrt{R},+1/\sqrt{R}\right\} $,
which makes $\mathbf{\Psi}$ a Rademacher matrix with unit column
norm. To obtain the estimate $\hat{\mathbf{b}}$ for the original
$\mathbf{b}$ from the noisy observation $\mathbf{v}$ we apply the
approximate message passing algorithm, as outlined in Algorithm \ref{alg: AMP-with-discrete}.
For a comparison we also apply to the same problem the algorithms
for the BPDN problem from \cite{Berg2007,Berg2008}, named SPGL1-BPDN,
and from \cite{Donoho2010}, named AMP-BPDN. Unlike the proposed AMP
algorithm the SPGL1-BPDN and the AMP-BPDN algorithm require the parameter
$\gamma$ as input, which we choose as $\gamma=\sqrt{\lceil W/R\rceil}\cdot\sigma\cdot\sqrt{R}\cdot\sqrt{1+\sqrt{2}/\sqrt{R}}$.
In contrast to our proposed algorithm, which always returns an estimate
for $\mathbf{b}$ in $\left\{ -1,0+1\right\} ^{W}$, the SPGL1-BPDN
and the AMP-BPDN algorithm return estimates in $\mathbb{R}^{W}$.
For this reason one can threshold the $\hat{\mathbf{b}}$ obtained
from these two algorithms by $\alpha$ and take as the output $\mathrm{sign}\left(\hat{b}_{w}\right)$,
where $\left|\hat{b}_{w}\right|\ge\alpha$, and zero otherwise. Alternatively,
one can search for the $K$ entries in $\hat{\mathbf{b}}$ with the
largest magnitude and set them depending on their sign to $\pm1$
and all other entries to zero.

In Figure \ref{fig:Fig4PaperRademacherDetectionErrorProbVsVariance_W=00003D512_R=00003D205_K=00003D20}
the detection error rate, i.e., $\mathrm{P}\left(\hat{b}_{w}\ne b_{w}\right)$,
is plotted for AMP-BPDN, SPGL1-BPDN, and the algorithm proposed in
this paper versus the noise variance. Our simulations indicate that
the detection error rate of the SPGL1-BPDN and AMP-BPDN algorithm
are strongly dependent on the chosen threshold $\alpha$ and that
the decoding rule which simply chooses the $K$ largest entries of
$\hat{\mathbf{b}}$ achieves the best performance for both of these
algorithms (this version is plotted in Figure \ref{fig:Fig4PaperRademacherDetectionErrorProbVsVariance_W=00003D512_R=00003D205_K=00003D20}).
However, even in this case the detection error probability performance
of the SPGL1-BPDN and the AMP-BPDN algorithm is approximately 6dB
respectively 20dB worse than that of the proposed AMP algorithm with
discrete prior.

\subsection{Random Demodulator}

\begin{figure}[b]
\centering \includegraphics[width=1\columnwidth]{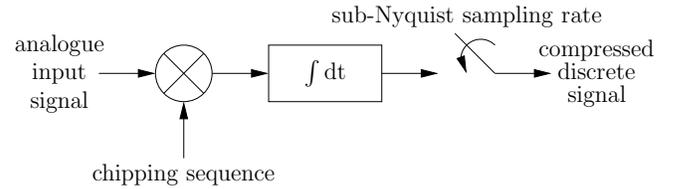}

\caption{Structure of the random demodulator as discussed in \cite{Tropp2010}}

\label{fig:randomDemod} 
\end{figure}

\label{sec:Random Demodulator} In this section we consider discrete
multi-tone signals, which occur, for example, in orthogonal frequency
division multiplex systems, in combination with the random demodulator.
For this class, it is shown in \cite{Tropp2010} how the operation
of the random demodulator, which is depicted in Figure \ref{fig:randomDemod},
on the analogue signal can be described equivalently by a time-discrete
representation. 

Let $\mathbf{F}\in\mathbb{C}^{W\times W}$ be a discrete Fourier transform
matrix and $\mathbf{b}\in\left\{ -1,0,+1\right\} ^{W}$ a $K$-sparse
vector. A time discrete representation of the analogue multi-tone
signal is then given by $\mathbf{x}=\mathbf{F}\mathbf{b}$. The multiplication
operation of the signal $\mathbf{x}$ with the chipping sequence and
the integrate and dump operation of the random demodulator are described
by the $W\times W$ diagonal matrix $\mathbf{D}$ and by $\mathbf{H}\in\left\{ 0,1\right\} ^{R'\times W}$,
respectively. $\mathbf{H}$ shall have $W/R'$ consecutive ones in
the $r$-th row starting from column $rW/R'+1$, where $r=0,\ldots,R'-1$.
As proposed in \cite{Tropp2010} we allow $\mathbf{H}$ to have fractional
elements in some of its columns when R does not divide $W$. To summarize,
the sparse signal $\mathbf{b}$ is observed by the random demodulator
through the random demodulator matrix $\mathbf{\Psi}'=\mathbf{HDF}\in\mathbb{C}^{R'\times W}$.
The sign of the non-zero entries of $\mathbf{D}$ on the main diagonal
is chosen independently at random. However, we choose the magnitude
of these entries here such that the columns of $\mathbf{\Psi'}$ have
unit norm.

As in Section \ref{sec:Rademacher Measurement Matrix} we aim here
to recover $\mathbf{b}$ from the noisy samples $\mathbf{v}$ given
in \eqref{eq:compressedNoisyMeasurement}, where however the real
valued measurement matrix 
\begin{equation}
\mathbf{\Psi}=\begin{bmatrix}\mathrm{Re}\left\{ \mathbf{\Psi}'\right\} \\
\mathrm{Im}\left\{ \mathbf{\Psi}'\right\} 
\end{bmatrix}\label{eq:componentWiseMeasurementMatrix}
\end{equation}
 depends now on the random demodulator matrix $\mathbf{\Psi}'$. For
our numerical experiments with the measurement matrix given by \eqref{eq:componentWiseMeasurementMatrix},
we set $W=512$, $R=204$, $R'=102$, and $K=20$. For this scenario
the detection error rate, i.e., $\mathrm{P}\left(\hat{b}_{w}\ne b_{w}\right)$,
is also plotted in Figure \ref{fig:Fig4PaperRademacherDetectionErrorProbVsVariance_W=00003D512_R=00003D205_K=00003D20}
versus the noise variance. The performance of all three algorithms
is very similar to the case in Section \ref{sec:Rademacher Measurement Matrix}
with a Rademacher measurement matrix. The performance gap between
the SPGL1-BPDN, the AMP-BPDN, and the proposed AMP algorithm with
discrete prior is here also approximately 6dB and 20dB wide.

\section{Conclusions}

\label{sec:Conclusions} In this paper we have developed a novel reconstruction
algorithm for compressive sampling problems, which applies the prior
knowledge that the entries of the original signal vector belong to
a finite alphabet. Our simulation results show for this algorithm
significant performance gains over existing reconstruction algorithms
for the BPDN problem. In future work we hope to further simplify the
message passing update equations of the proposed algorithm and thus
further reduce its complexity.

\begin{algorithm}[t]
\footnotesize
\begin{itemize}
\item \textbf{Input}: Measurement matrix $\mathbf{\Psi}$, observation $\mathbf{v}$,
noise variance $\sigma^{2}$, alphabet $\mathcal{A}_{0}$, prior probabilites
$\pi_{0},\pi_{1},\ldots,\pi_{S}$, number of iterations $T$
\item \textbf{Output}: signal estimate $\mathbf{\hat{b}}$\end{itemize}
\begin{enumerate}
\item Initialize: $t=1$, and $l_{w\to r}^{(0)}(a_{s})=\pi_{s}$, for $w\in[W]$,
$r\in[R]$, $s\in[S]$.
\item Calculate the mean and variance of the variable-to-factor messages:\vspace{-0.3cm}

\begin{align*}
\xi_{w\to\cdot}^{\left(t\right)}= & \sum_{s=0}^{S}a_{s}\exp l_{w\to\cdot}^{\left(t-1\right)}\left(a_{s}\right)\\
\tau_{w\to\cdot}^{\left(t\right)}= & \sum_{s=0}^{S}a_{s}^{2}\exp l_{w\to\cdot}^{\left(t-1\right)}\left(a_{s}\right)-\left(\xi_{w\to\cdot}^{\left(t\right)}\right)^{2}
\end{align*}

\item Approximate the mean and variance of $\prod_{\rho\neq\cdot}\hat{\nu}_{\rho\to w}^{\left(t\right)}\left(b_{w}\right)$:\vspace{-0.3cm}

\begin{eqnarray*}
\mu_{w\to\cdot}^{\left(t\right)} & = & \sum_{\rho\neq\cdot}\psi_{\rho,w}\left(v_{\rho}-\sum_{\omega\neq w}\psi_{\rho,\omega}\xi_{\omega\to\rho}^{\left(t\right)}\right)\\
\eta_{w\to\cdot}^{\left(t\right)} & = & \frac{1}{R}\sum_{\omega\neq w}\tau_{\omega\to\cdot}^{\left(t\right)}+\sigma^{2}
\end{eqnarray*}

\item Incorporate prior and normalize:\vspace{-0.3cm}
\begin{eqnarray*}
\bar{l}_{w\to\cdot}^{\left(t\right)}\left(a_{s}\right) & = & \log\pi_{s}-\frac{\left(\mu_{w\to\cdot}^{\left(t\right)}-a_{s}\right)^{2}}{2\eta_{w\to\cdot}^{\left(t\right)}}\\
l_{w\to\cdot}^{\left(t\right)}\left(a_{s}\right) & = & -\log\left(\sum_{s'=0}^{S}\exp\bar{l}_{w\to.}^{\left(t\right)}\left(a_{s'}\right)\right)+\bar{l}_{w\to.}^{\left(t\right)}\left(a_{s}\right)
\end{eqnarray*}

Set $t\leftarrow t+1$, and repeat (2)-(4) until stopping criterion
holds. 

\item Return $\hat{\mathbf{b}}=\left[\hat{b}_{0},\ldots,\hat{b}_{W-1}\right]^{\mathrm{T}},$
where
\begin{eqnarray*}
\hat{b}_{w} & = & \arg\max_{a\in\mathcal{A}_{0}}l_{w\to\textrm{ø}}^{\left(t\right)}\left(a\right).
\end{eqnarray*}

\end{enumerate}
\caption{\label{alg: AMP-with-discrete}AMP with discrete prior}
\end{algorithm}

\footnotesize\bibliographystyle{IEEEbib}
\bibliography{ReferencesPaperOnCompressiveSampling}

\end{document}